\documentclass[12pt]{article}

\begin{document}

\begin{center}
{\bf \Large Top 10 Problems on Massive Stars}\vspace{1cm}\\
{\bf C\'assio L. Barbosa (IAG-USP) \& Donald Figer (STScI)}\\
cassio@astro.iag.usp.br \hspace{2cm} figer@stsci.edu\\
(Organizers)\vspace{1cm}
\end{center}

{\bf \large A foreword}\vspace{0.5cm}

Massive stars are rare and fascinating. While heavily obscured in the
earliest stages of their lives, they evolve to become the brightest sources in
their host galaxies when they die as supernovae. Their lifetimes are short,
but the effects of massive stars in the galaxies are dramatic, altering their
environments on both global and local scales. 

Ten years ago, Dr. Andre Maeder compiled a list of 10 Most Topical
Stellar Problems\footnote{Frontiers of Space and Ground-Based Astronomy: The
Astrophysics of the 21st century, Dordrecht, Boston: Kluwer Academic
Publishers, c1994, edited by W. Wamsteker, Malcolm. S. Longair, and Y. Kondo.
Astrophysics and Space Science Library, Vol. 187, p.177)} that became a
reference for many astronomers, especially those new to the field. Since 
then, we have witnessed the birth of new ideas, and rapid improvement of
theories and technologies that have had a profound impact on the field. 
We believe that it is time again to ask, what
are the top 10 most interesting problems regarding massive stars? 

To this end, we have asked a number of researchers in the field to compile
their lists of the top 10 problems in the field of massive stars.
This paper is a compilation of these lists. We attempted to survey 
observers and theorists and those studying all evolutionary stages in the
lives of massive stars. Each list reflects the proposer's
personal point of view, but hopefully, this compilation will
motivate new young astronomers and begin a new series of discussion.

\newpage

\noindent {\bf Dr. Andre Maeder}

\begin{enumerate}

\item In the  formation of massive stars, what is  the relative role
of accretion and collisions?
\item How massive stars do form at metallicity Z$\sim$0?  What is the IMF,
upper, lower mass limits and the  fraction of binaries at  Z$\sim$0?
\item What are the distributions of rotational velocities at lower Z?
\item What is the role of magnetic field in formation and evolution of
massive stars?
\item About the mass loss rates, what are they at different metallicities 
for all kinds of stars, OB stars, supergiants, WR stars?
\item What is the physics and role of internal mixing from MS to
supernova explosions, for different masses and Z?
\item What is the exact role of binarity and mass transfer for the 
value of the chemical yields and the populations of massive stars?
\item What are  the initial mass limits for the formation of neutron stars 
and black holes? How do these limits change with Z?
\item How to explain the rotation periods of young pulsars  and the
higher rotation necessary for the collapsar model?
\item Where is exactly the mass cut between what is ejected in
supernova and what is captured in the remnants? How does it change with Z? 

\end{enumerate}

\newpage

\noindent {\bf Dr. John Bally}

\begin{enumerate}

\item ACCRETION OR MERGERS?
   Do massive stars form by direct accretion from a cloud or disk
   or by merging with other stars?  Do both mechanisms operate?
   Which one dominates?

\item A FORMATION SEQUENCE:
   Is there an empirical evolutionary sequence for massive stars
   which is analogous to the Class 0, Class I, Class II,
   Class III sequence?

\item THE MASSES OF MASSIVE STARS:
   What determines the mass of a massive star?  What determines
   the IMF in a massive cluster?  -  What determines the upper-mass
   limit of massive stars?  How does this mass limit depend on
   metallicity, stellar spin, and other
   environmental or intrinsic properties?

\item SUPERMASSIVE OBJECTS:
   What happens to a collapsing cloud that attempts to form a
   supermassive star?

\item THE FIRST STARS:
   What are the properties of zero-metallicity population III
   massive star forming regions?  What is expected from theory?
   What can we observe?

\item STARS IN GALACTIC NUCLEI:
   How do massive stars form in galactic nuclei?   How did the
   Sgr A cluster of massive stars form and evolve?

\item CONNECTIONS TO AGN:
   What are the relationship between massive star formation,
   super star cluster formation, globular cluster formation, and
   the birth of black holes, super-massive black holes, and the
   powering of AGN activity?

\item OUTFLOWS:
   How are protostellar outflows produces by massive stars produced?
   What processes are similar (different) to those in low-mass stars?

\item MAGNETIC FIELDS:
   What are the roles of magnetic fields in massive star formation?

\item MULTIPLES: How do massive star binaries and multiples form?
   Is the process similar to, or different from the formation of
   low-mass multiples?

\end{enumerate}

\newpage

\noindent {\bf Dr. Ed Churchwell}

\begin{enumerate}

\item Do massive stars form primarily via accretion or by a combination of early
accretion followed by mergers of low to intermediate mass protostars?

\item If by accretion then:

\begin{itemize}
\item[a)] What is the mechanism that drives bipolar outflow?
\item[b)] What is the relationship between accretion, accretion disks, and
bipolar outflows?
\item[c)] What is the role of gravity, radiation and stellar winds in
protostars?
\end{itemize}

\item If by mergers then:

\begin{itemize}
\item[a)] What are the unique observational characteristics?
\item[b)] How do we interpret the known massive outflows associated with some
massive star formation regions?
\end{itemize}

\item Can we identify and understand the physical processes that determine the
early evolutionary sequence of massive stars? Is the sequence: prestellar cores
$\rightarrow$ hot cores $\rightarrow$ hypercompact HIIs $\rightarrow$
ultracompact HIIs $\rightarrow$ compact HIIs and what changes occur with the
central star through these phases?

\item Since most stars form as members of binary or multiple systems, this has
to be included in any successful theory of star formation. Binary and multiple
systems are natural consequences of the merger hypothesys but not so clear for
the accretion hypothesis. Understanding the physics of multiple star system
formation is a fundamental goal.

\item The distribution of stellar masses (i.e. the IMF) in young clusters
appears to be essentially constant in the Galaxy despite formation in a wide
range physical conditions (velocity fields, density, chemical composition, tidal
forces, etc.). It is not understood what controls the IMF and until we do, we
cannot claim that star formation is understood.

\item What is the role of magnetic fields in star formation? It is generally
believed that star formation occurs by accretion and that magnetic fields
somehow control the flow of matter into the associated bipolar outflows.
However, only nature understands how this actually works.

\item Massive star formation occurs in highly obscured molecular clouds, so it
has been very difficult to determine the stellar census of very young massive
star formation regions. Deep MIR and NIR imaging and spectroscopic observations
of these regions need to be obtained to reveal the location and relative
numbers of stars that accompany the massive protostar that gives rise to
hypercompact and ultracompact HII regions. Is the IMF in these very young
regions the same as in older more evolved clusters?

\item Molecular line surveys of the Galaxy shows that the Milk Way has a
prominent molecular ring with a galactocentric radius of $\sim$5 kpc and a
central bar. We know almost nothing about the stellar content and rate of star
formation in these prominent components of our galaxy. This is a problem that
needs serious attention.

\item What is the role of dust in the process of star formation? Is dust
composition, size distribution, and charge critical to the process?

\item Is it possible for massive protostars to form planets? Are their accretion
disks destroyed before planet formation can occur? Do the conditions in their
accretion disks permit planet formation?

\item What is the origin of the mass in the outflows associated with massive
protostars?

\item How did the first stars in the universe form in the absence of metal
coolants? What were the properties of these stars?

\end{enumerate}

\newpage

\noindent {\bf Dr. Peter Conti}

Stellar astrophysics involves the evolution of stars, their location,
impact on their surroundings, and their role in the Universe.

Several of the most perplexing problems of the life of stars are
concerned with their birth.  There has been considerable progress over
the past few decades in modeling their normal lifetimes and understanding
the death processes, but the initial formation has typically been hidden
from view by the presence of the natal clouds of dust and gas.  As new
wavelengths have opened to exploitation with improving telescopes and
instrumentation we are finally beginning to be able to observe the birth
itself.

\begin{enumerate}

\item Most OB stars form in clusters and associations from Giant Molecular
Clouds but do all of them?  Some stars are found well away from any known
large stellar groupings. Have they escaped (the available time scale is
short) or were they themselves merely the brightest members of otherwise
rather skimpy small associations lower down on the molecular cloud IMF?

\item Does an individual star grow in mass from stellar accretion via a
natal disc, or from mergers of smaller objects in a tight cluster
environment? Could both of these processes play a role?  Are there
differences in the birth mode between single and close (or more distant)
binary systems?

\item The terms ``hot cores" and UCHII regions refer to some of the earliest
stages of massive star birth. The former term refers to protostellar gas
clouds that are already detectable from their thermal IR radiation; the
latter term has to do with an ionized hydrogen region surrounding an
already existing star hot enough to release prodigious amounts of Lyman
continuum photons. Are the ``hot cores" not quite hot enough (yet?) to
generate UCHII regions or are they already doing so but their radiation
is still buried within the natal clouds? Is there any distinction between
hot cores and UCHII regions excited by close binaries or by single stars?

\item Most massive stars are born in giant molecular clouds.  What is the
effect of this dense stellar and gaseous dusty environment on the very
early evolution of an individual object?

There has been dramatic improvement of our understanding of the evolution
of massive stars during the last few decades. The following unresolved
issues come to mind:

\item What is the minimum mass above which a massive star will become a
Wolf-Rayet type? How does this depend on the metallicity? In the close
binary case, this number might well be lower, but by how much? It is even
correct to talk about a minimum mass?

\item What is the minimum mass above which a massive star will end its life
as a black hole, rather than as a neutron star? How does this depend on
the metallicity, or the close binary nature? Is it even correct to talk
about a minimum mass?  Is this number related to the formation of W-R
stars? Do all massive stars end their lives as supernovae or do some of
them (the most massive?) turn directly into black holes?

\item What is the role of rotation and chemical mixing on the evolution of
massive stars, both OB and W-R types? The importance of stellar winds in
addressing this problem has been realized and its connection to rotation
is in the beginning stages of solution. How do these physical processes
couple? What role do close binaries play?

The location of massive stars in our Galaxy shows them to be closely
confined to the galactic plane, so we are clearly in a spiral system.

\item What is the structure of the spiral arms? It is believed that we are
in a multiple arm system given the longitude locations of OB stars near
the sun. But does this carry around to sites beyond the galactic center,
or are we being misled by the local view? How would our Galaxy appear
optically (in the UV?) if viewed from well above (or below) the plane?

During their lifetimes massive stars eject ionizing photons, considerable
wind energy, and processed nuclear material into their immediate
environments. Their death as supernovae causes another spike of energy
and a substantial ejection of ``heavy" elements into the interstellar
medium.

\item How does the importance of these processes depend on the initial
metallicity? What role do close binaries play in modifying these events?

As we inquire into how the Universe has evolved, we realize that ``In the
beginning" there were massive stars. Much larger telescopes and greatly
improved instrumentation are now beginning to probe the very youngest
galaxies and their constituent stars. These will be predominantly the OB
types (although that classification will need to be modified in the no
``metals" case).

\item How do massive stars evolve in the limit of no ``metals" during the
earliest stages of the evolution of the Universe? Modeling using both
rotation and stellar winds will need to be made.  Do either of these
processes play a role in causing internal mixing and could that process
modify how an individual first generation massive star evolves?

\end{enumerate}

\newpage

\noindent {\bf Dr Philippe Eenens}

Here are a few questions which puzzle me:

\begin{enumerate}

\item Massive star birth;

\item Wind velocity law of WR stars;

\item Departure of WR stars from spherical winds;

\item Rotation rates of WR stars;

\item Evolutionary links between B[e], LBV and WR stars;

\item Presence of dust in hot winds;

\item Evolution of WR into supernova;

\end{enumerate}

\newpage

\noindent {\bf Dr. Neil Evans II}

\begin{enumerate}

\item Primordial Star Formation, with only H$_{2}$ as a coolant.

\item How much metallicity is needed before the nature of star formation changes
   enough to permit a fuller range of masses?

\item What drives starbursts in high-z galaxies and what are properties
   of the stars (IMFs, etc.)?

\item How similar are the conditions in dusty starbursts to those in
   massive star forming regions in our Galaxy?

\item What controls the IMF in current (Galactic) star formation?

\item Can we actually see the fragmentation in massive dense cores
   in an unbiased tracer?

\item What chemical changes go on and how do they affect observed line profiles?

\item How are the dust properties changing during massive star formation,
   and how do those changes affect observed emission from dust?

\item Can we trace the dynamical process of collapse and fragmentation with
   any molecular line tracer?

\item What are the feedback effects of massive stars on the formation
    of other stars?

\end{enumerate}

\newpage

\noindent {\bf Dr. Henny Lamers}

In my order of priority:

\begin{enumerate}

\item How do massive stars form?

Co-agulation requires very fine-tuning of the
conditions and time scales. I am doubtful that it will work.

\item What causes the high mass loss rates of WR stars? 

The present radiation
driven wind models start at the photosphere. but there must be a very deep
subsonic moving (!) region between the hydrostatic core (about 1 R$_{\odot}$!)
to the photosphere (about 10 -- 20 R$_{\odot}$). The mass loss rate is
determined somewhere deep down, possibly just above the hydrostatic core.
(ps: Nugis and Lamers are trying to model this)

\item How does the mass loss rate vary with metallicity at low metallicities?
This is important for the evolution of the first generation stars.

\item What causes the outbursts of LBVs?

They are close to their Eddington limit, even deep inside the star so they are
probably marginally stable. Any internal disturbence might have a large effect.
But what is this disturbence? What sets the time scale?

\item What is the role of rotation of massive stars?

The final fate will depend crucially on the rotation and on the transport of
angular momentum. (I have tried to derive this observationally by studying
the abundances of WR and LBV nebulae, and conclude that rotation induced mixing
must occur on a timescale that is several times the MS-lifetime 
(Lamers, Nota, Panagia,  et al. ApJ few years ago).

\item Do all massive stars form in clusters?

In our galaxy: probably yes. But there may be special environments where this
is  not the case. e.g. near the core of M51 (Lamers et al, 2003,ApJ) Are there
other cases where this happens?

\item What sets the upper mass limit for star formation?
     
I always thought that is is the Eddington limit, but Norbert Langer tells me
that is not the case. A very massive star can survive by becoming convective
and thus reducing the radiative flux and the radiation pressure (this also
happens in very luminous WR stars). This question may be related to point 1
(formation)

\item Is the IMF for massive stars the same everywhere? (related to 1 and 7)

\end{enumerate}

\newpage

\noindent {\bf Dr. Norbert Langer}

\begin{enumerate}

\item How do the most massive stars form? Which are the highest masses made?
    Does the formation process depend on metallicity or environment?

\item How do massive stars explode when they form a neutron star?
    Does the "prompt explosion mechanism"
    apply to the lowest mass iron cores; perhaps in close binaries?
    Is it the neutrinos that are responsible for most supernovae in
    the "delayed explosion mechanism"? What produces the neutron star
    kicks?

\item Do massive stars explode when they form a black hole? Which fraction of
    them creates gamma-ray bursts, which hypernovae? Is there a specific
    role for massive stars in close binaries? Do black holes receive a kick
    when they are born?

\item What is the role of magnetic fields in the interior of massive stars?
    Are the torques exerted by them strong enough to slow down the core so
    that we can understand the spin rates of young pulsars? But if so, how
    do we get the rapidly rotating cores required for the collapsar model
    of gamma-ray bursts?

\item Which is the site of the r-process. Massive stars, for sure but which
    and where? And how many r-process components exist?

\item In a mass transferring massive binary, which fraction of the overflowing
    matter can be retained by the mass receiving star? This question is
    relevant to all post-mass transfer binaries: massive Algols, Wolf-Rayet
    binaries, X-ray binaries, double-neutron star binaries, Type Ib/c
    supernovae, gamma-ray bursts (long and short),...
    It is also strongly connected to the star formation process: if disk
    accretion is involved, it may work the same way in newly born stars as
    in "rejuvenated" ones in close binaries.

\item What drives red supergiant mass loss? How big are the mass loss rates?
    How does the mass loss rate depend on metallicity, or envelope chemical
    composition?

\item What causes LBV giant eruptions? Why are all LBV nebulae bipolar?
    What is the metallicity dependence of
    the LBV eruptions? I.e., at very low metallicity, do the most massive
    single stars go through an LBV stage, thereby lose most of their envelope
    form Wolf-Rayet stars? If not, there may be no other way to for WR
    stars, with two consequences: single stars could not form gamma-ray
    bursts from collapsars, and stars above 60...100 solar masses would
    not form iron cores but explode as pair creation supernovae during
    oxygen burning.

\item What is the origin of the Ultraluminous X-ray Sources? Are there black
    holes of 100 Msun and more in binary systems? Is there a continuous
    mass spectrum of black holes, from stellar (2...15 M$_{\odot}$) to
    supermassive (...10$^{8}$ M$_{\odot}$) ones?

\item How do supernova ejecta merge with the interstellar medium?
     What is the role in this of the pre-supernova shaping of the
     circumstellar medium from the supernova progenitor's winds and
     photons? What is the role of supernovae for star formation?
     How do the hot ejecta mix with molecular clouds? Which are
     particles are accelerated to form the cosmic rays?

\end{enumerate}

\newpage

\noindent {\bf Dr. Anthony Moffat}

\begin{enumerate}

\item Need for a self-consistent, complete model of massive-star evolution
(e.g. without having to assume a beta-law for the wind!)

\item What is the birth scenario of massive stars?

\item What is the upper mass-limit of star formation?

\item Is the IMF the same for single stars as those in binaries?

\item What is the initial binary frequency;  does it vary with Z, mass, ...?

\item Under what conditions does RLOF take place and how is it important?

\item What is the mechanism behind LBV eruptions?

\item What happens at Z = 0 (pop III), where all stars are probably massive?

\item How does dust form?

\item Do massive stars produce GRBs (short ones from merging of NS/NS or NS/BH,
long ones from collapsing rapid rotators?)?

\item Do WR stars explode as SN Ib/c?

\item What is the structure of WR winds in the optically thick zone and what
are the true hydrostatic radii?

\item Are all stars formed in clusters/groups; origin of isolated OB/WR stars?

\end{enumerate}

\newpage

\noindent {\bf Dr. Jonathan Tan}

\begin{enumerate}

\item What are the lifetimes and formation mechanisms of Giant Molecular
Clouds (GMCs), the hosts of all present-day massive star formation?
(i.e. how long do they exist as coherent structures before being
dispersed, how does this compare with the timescale for the
dissipation of their turbulence, and how important is continued energy
injection from internal star formation or external perturbations?)

Observationally, the mass fraction of a galaxy's interstellar medium
(ISM) that is associated with GMCs (and associated atomic halos)
approximately tells us the relative amount of time gas spends in bound
and unbound states. Together with other data, e.g. cloud angular
momenta, this helps constrain formation scenarios.  On the theoretical
side, numerical simulations of magneto-hydro-dynamic turbulence
indicate short decay times of just a few million years, but it is not
yet clear if these models capture all the physics of real clouds.

\item What, if any, is/are the triggers for star cluster formation from
highly localized regions of GMCs, and how does this answer relate to
trends in total galactic-scale star formation, such as the
Schmidt-Kennicutt relations?

This question will probably be answered observationally via a study of
the regions that are precursors to those of active massive star
formation. These are best identified via mid-infrared absorption
(e.g. the dark clouds seen by the ISO and MSX satellites). Then an
association with spiral arms, cloud collisions, supernova remnants,
etc. can be searched for.

\item Does massive star formation (or indeed any kind of star formation
in star clusters) proceed in a manner best modeled as the collapse of
quasi-equilibrium structures (``cores'') or as competitive Bondi-Hoyle
accretion? 

Apparently coherent, equilibrium cores are commonly observed, with a
mass function that is apparently quite similar to the stellar initial
mass function (IMF). These observational studies need to be extended
to the more distant and denser regions characteristic of high-mass
star formation. Reconciling the simulations of a turbulent ISM with
the apparent existence of equilibrium cores is an important goal.

\item What sets the shape and upper limit of the stellar IMF, and does
this vary for different initial conditions of the interstellar medium?

There appears to be a physical process that limits the mass of
Galactic stars (the IMF appears to be a power law truncated above
about $\sim 100-200 M_\odot$), but it is not known whether this is due
to the action of feedback during the formation process, is due to
dynamical processes in the forming star clusters, or is due to some
instability in the structure of very massive stars that initiates
rapid mass loss.  The power law shape may be already present in cores
(see Q3) or may result independently. Massive star formation is
observed to occur in a wide range of conditions, such as dense clumps
inside GMCs, the high pressure environment of our Galactic center, and
in massive, relatively low-metallicity clouds in dwarf galaxies such
as the Large Magellanic Cloud, and it is important to continue looking
for variations in the IMF from region to region. Surprisingly no
differences have been found in such studies. Within individual
protoclusters there is tentative evidence that massive stars are
biased to the more central regions: more observations, coupled with
detailed n-body modeling of specific clusters are needed to confirm this.

\item Are there disks around massive protostars, and, if so, how do their
properties (size, mass, etc.) compare to those around lower mass stars?

There have been several interesting observational claims for massive
protostellar disks, but so far none (in my opinion) are completely
unambiguous. Another question is if disks are present, then do they
maintain their orientation over many dynamical timescales - i.e. is
their orientation set by overall angular momentum of a coherent core
surrounding and feeding the disk - or does the orientation change
because the feeding is more erratic?

\item What determines the nature of massive star binaries? Do planets
form around massive stars?

Answering this question can help lead to a better understanding of the
star formation mechanism: do binaries and/or planetary companions form
from a common disk inside a core, by chance dynamical interactions
in a crowded star cluster, or by a combination of both mechanisms? As
observational techniques to find planets improve, it may become
possible to look at planet frequency over a wide range of stellar
masses, as has been done for protostellar disks in young star
clusters.

\item How are runaway OB stars created?

Again observational studies of individual systems are the best hope
for answering this question, with the leading theories being from
three body interactions in young star clusters, or from the
supernova explosion of a massive star with a massive companion.

\item What is the nature of outflows from massive protostars - in
particular, are they driven magnetocentrifugally (via disk winds or
X-winds) as is thought to be the case with low-mass protostellar
outflows, or does radiation pressure star to become more important?

Outflows around massive protostars may well confine hypercompact HII
regions, and VLBA studies of these sources (e.g. source ``I'' in Orion)
can help determine the properties of the outflow on scales of tens of
AU. Maser observations can also help determine the outflow properties,
although there remains the caveat of the precise relationship between
observed maser spot velocity and true flow velocity. Outflows from
massive protostars should initially have speeds of order 1000 km/s.

\item What feedback processes, if any, determine the efficiency of star
formation from both individual cores or protoclusters?

This is a very complicated problem, and for the moment is beyond an
accurate numerical study. Simple analytic and semi-analytic models may
give some guidance, but only if they are tied closely to observations
of young and forming star clusters.

\item How does including rotation affect protostellar structure and
evolution models?

By determining the size of the star, protostellar structure affects
the stellar luminosity and the strength of outflows from the inner
accretion disk. Stellar evolution models including rotation are now
available, and these need to be extended to include the protostellar
phase.

\end{enumerate}

\newpage

\noindent {\bf Dr. Nolan Walborn}

\noindent My Top (Ten) Eleven! Problems on Massive Stars

\begin{enumerate}

\item What are the physical parameters and generic relationships of the
several classes of optically observable, very young (near or on ZAMS)
massive stars?  Several categories and phenomena (not always in the same
objects) have been described, but there has been little or no directed,
systematic, quantitative analysis of them.  They include objects in 
dense nebular knots, with excessively strong Balmer lines (class Vb) or 
He II 4686 stronger than other He II lines (class Vz, inverse Of effect?), 
with weak UV wind profiles for their types, and subluminous (fainter than
class V) objects.

\item What are the true distances and luminosities of OB stars in the solar
neighborhood?  This question will be answered only by SIM and GAIA.

\item Which OB stars in the solar neighborhood and the Magellanic Clouds are
single objects and which are unresolved multiple systems?  With current
instrumentation, there is an unobservable gap between the
spectroscopic-binary limit of ~1 AU and the interferometric imaging
limits of, e.g., 25 AU in the Carina Nebula or 500 AU in the LMC (for 
components of similar brightness).

\item What are the true masses and the upper mass limit of OB stars?  
Progress is being made, but we need to have a definitive temperature 
scale and resolution of systematic differences between different methods 
of determining masses (spectroscopic/eclipsing binaries, atmospheric
spectral analysis, evolutionary models).

\item Do the statistics of CNO line strengths in OB supergiants (OBC vs.
morphologically normal vs. OBN classes) correspond to those of initial
main-sequence rotational velocities?  Are there systematic rotational
differences corresponding to relative surface CNO differences among
different clusters?

\item Are the winds of normal OB and WR stars spherically symmetric or not?
What is the physics of WR winds?

\item What are the physical parameters and generic relationships of the many
varieties of luminous, evolved, emission-line OB stars, i.e. the OB Zoo?
They include the following categories: WN-A (WN6-8L), O Iafpe and
Ofpe/WN9 (WN9-10ha), B Iape (WN11h), Iron (Fe II), and LBV.  Some
relationships have been established, e.g. Ofpe/WN9 and B Iape spectra
have been observed in the minimum state of certain LBVs, but the global
relationships are far from clear, and may well be functions of mass and
metallicity.

\item Which of the foregoing and other phenomena correspond to evolutionary
states of single massive stars, and which arise from close-binary
interactions?  Starting with Eta Carinae!

\item What are the OB progenitors of the multiplying varieties of
core-collapse SN?  And of GRB? 

\item Have we identified and correctly modeled the fundamental parameters
that determine massive stellar evolution in detail?  At the highest
level, they are now believed to be mass, metallicity, and initial
rotational velocity.  Are magnetic fields a significant variable or not?
What are the evolutionary tracks and sequences of spectral states
corresponding to all relevant combinations of these parameters?

\item And finally, we ultimately need physical models that reproduce observed 
spectra and luminosities from first principles!  Current models are replete 
with ad hoc parameters (e.g., convection, mixing, mass-loss rates) that are 
adjusted to apparently match observations, but there is no guarantee that 
such procedures are unique or correspond to real phenomena in nature.

\end{enumerate}

\newpage

\noindent {\bf Dr. Stan Woolsey}

\noindent Eight top issues:

\begin{enumerate}

\item How do massive stars die?

   After 50 years of research we still don't know. Do rotation
  and magnetic fields play a role or can neutrinos alone
  make a succesful supernova explosion? Even the non-rotating, non-magnetic
  calculations strain current codes and resources beyond credibility.

   The location of the mass cut influences the composition of the
    universe, which SN make black holes and which make neutron stars,
    and the nature of GRBs.

\item What is the evolution of angular momentum inside massive star?

    If the star rotates rigidly on the ZAMS, what is the j of the final
collapsed remnant? How does rotational mixing affect the evolution and the
abundances seen at the surface?

\item What is the mass loss rate as a function of metallicity, mass, and
evolutionary stage

     What is the mass of the star when it dies and how does mass loss
   affect nucleosynthesis?
    What are the mass loss rates of WR stars of various mass and initial
   metallicities?

\item How are massive stars formed?

   Is their birth fundamentally different from low mass stars. What is
the metallicity dependence of the star formation process and a related questi

\item How has the IMF evolved with metallicity in the universe. What were
the masses of typical stars with Z = 0? What was the heaviest star made? What
its mass when it died?

\item How to treat convectiive boundries and semiconvective regions

     What is the overshoot length as a funtion of e.g., the pressure scale
   height or the entropy contrast? What is the semiconvective diffusion
   coefficient?

\item How has membership in binaries affected the evolution and the
properties of the star at death?

\item What nucleosynthesis has each massive star of a given Z and mass
contributed?

\end{enumerate}
      
\end{document}